\documentclass[aps,twocolumn,prb]{revtex4-1}

\usepackage{multirow}
\usepackage{amsmath}
\usepackage{amsfonts}
\usepackage{graphicx}
\usepackage{color}
\usepackage{bm}
\usepackage{dcolumn}
\draft 

\begin{document}

\title{Assessing interface coupling in exchange-biased systems via in-field interaction plots \\
       (\rm Journal of Magnetism and Magnetic Materials, \textcolor[rgb]{0.00,0.00,1.00}{doi.org/10.1016/j.jmmm.2019.166061})   }

\author{ J. Geshev,$^{1}$ L. L. Bianchi,$^1$ R. F. Lopes,$^{1,2}$ J. L. Salazar Cuaila,$^1$ and A. Harres$^3$}
\affiliation{\small \\~\\
\mbox{$^1$Instituto de F\'{\i}sica, URFGS, Porto Alegre, 91501-970 Rio Grande do Sul, Brazil} \\
\mbox{$^2$Instituto Federal Sul-rio-grandense, C\^ampus Sapiranga, Sapiranga, 93804-870 Rio Grande do Sul, Brazil} \\
\mbox{$^3$Departamento de F\'{\i}sica, UFSM, Santa Maria, 97105-900 Rio Grande do Sul, Brazil} \\ }

\date{\today}

\begin{abstract}
An in-field interaction plot, $\delta M_{\rm R}$, has been recently introduced, presenting important advantages over the classical remanence plots. Here a general $\delta M_{\rm R}$ is proposed, allowing to assess interactions even in systems with shifted and asymmetric major loops. To construct such a plot, a recoil loop (which incorporates a first-order reversal curve, FORC) and the position of the center of the major loop are only needed. Applying the method on exchange-biased Co/IrMn bilayer gives two types of $\delta M_{\rm R}$ obtained for measuring field either parallel or antiparallel to the exchange-bias direction. This provides valuable information on the reversal mechanism and allows distinguishing between effects coming from coupling into the ferromagnet (Co) and those stemming from interactions in its interface with the antiferromagnet (IrMn). The essentially nonzero general $\delta M_{\rm R}$ plot obtained from the major loop revealed to be a precise measure of the reversal asymmetry. The technique can readily be adjusted for use in other scientific fields where hysteresis is observed. We provide free software which generates such $\delta M_{\rm R}$ plot(s).
\end{abstract}


\maketitle

\section{Introduction}
Wohlfarth pointed out a simple relation\cite{Wohlfarth-1958} between remanence curves of systems with symmetric major magnetization $M$, versus magnetic field $H$, hysteresis loops. These are the isothermal remanent magnetization curve $M_{\rm r}(H)$, which represents the remanence obtained by the application and removal of a positive field on an initially demagnetized sample, and the DC demagnetization curve $M_{\rm d}(H)$, i.e., the remanence resultant from the application of a negative field to a sample initially at saturation remanence. The Wohlfarth's relation
\begin{equation}
\label{Wohlfarth}
M_{\rm d}(H)= 2M_{\rm r}(H) - M_{\rm r}(\infty)
\end{equation}
should be valid for non-interacting uniaxial-anisotropy systems no matter whether the magnetization reversal occurs via domain nucleation followed by domain-wall motion or coherent rotation.

Nonzero $\delta M(H)\,$$=$$\,2M_{\rm r}(H)\,$$-$$\,M_{\rm d}(H)\,$$-$$\,M_{\rm r}(\infty)$ values of, e.g., initially thermally or AC demagnetized samples, are ascribed to magnetic interactions; positive values are normally attributed to exchange-like coupling favoring a ferromagnetic state and negative values are associated with dipolar-like interactions stabilizing the demagnetized state.\cite{Gaunt,Kelly,Bissel,Mayo} Non-interacting cubic-anisotropy systems present intrinsically positive $\delta M$ plots.\cite{Rems-cubic-A}

In exchange bias (EB) systems with shifted (by the so-called EB-field, $H_{\rm eb}$) and often asymmetric major hysteresis loops, $\delta M$ plots cannot be used in their classical forms. Even though this technique has been adapted to biased systems,\cite{Harres-JAP-2013} it still requires demagnetization. Interaction plots based on initial magnetization and hysteresis curves\cite{dMa-A,Thamm-JMMM-1996} are easy to obtain and present characteristics very similar to those of the remanence ones. Nevertheless, these still require an initially demagnetized state.

Generalized and/or integral $\delta M$ plots and functions, obtained with the help of first-order reversal curves (FORCs),\cite{FORC} have also been proposed.\cite{Bissell-1994, Buehler-Mayergoyz-1996,Stancu-FORCs,Bissell-2000} Methods based on FORCs and also on second-order reversal curves (SORCs) and remanent SORCs\cite{Stancu-JAP-2006, Bodale-IEEE-2011} have been used to study magnetization reversal mechanisms and interactions as well. Pike {\it et\,al.}\cite{Pike-JAP-1999} have claimed that FORC diagrams give more precise information on magnetic interactions than the $\delta M$ plots. FORCs have also been used in studies of magnetic interactions in EB systems.\cite{Cornejo-2010,Khanal-Gallardo,Toro-CM-2017} However, the greatly-increased amount of FORC and SORC data as compared to those of the remanence plots, together with the complexity of their analyzes, could make their interpretation rather difficult, particularly true when magnetic interactions are present and the Preisach-like interpretation is not applicable. The technique is often considered as a magnetic fingerprint and not a method that provides quantitative information.\cite{Ruta-2017,Goering-2019}

Recently, a relation analogous to that of Wohlfarth but between in-field magnetization curves has been derived\cite{dMr-2018} for systems with symmetric major hysteresis loops,
\begin{equation}
\label{Rec-Sym}
M_{\rm rec}(H) = 2\overline{M}_{\rm hys}(H) - M_{\rm sym}(H).
\end{equation}
Here $\overline{M}_{\rm hys}$$\,=\,$$\frac{1}{2}(M_{\rm dsc}$$\,+\,$$M_{\rm asc})$, being $M_{\rm dsc}(H)$ and $M_{\rm asc}(H)$ the descending and ascending branches of the major loop, and $M_{\rm sym}(H)$ the curve symmetric, in respect to the origin of the coordinate system, to the extended recoil curve $M_{\rm R}(H)$;\cite{dMr-2018} note that the latter also represents a FORC with reversal field $H_{\rm R}$. Based on Eq.\,\ref{Rec-Sym}, an in-field interaction plot has been introduced,
\begin{equation}
\label{dMg}
\delta M_{\rm R}(H) = M_{\rm R}(H) + M_{\rm sym}(H) - 2\overline{M}_{\rm hys}(H).
\end{equation}
It is acquired in an easier and faster manner than $\delta M$ and does not demand demagnetization, significantly simplifying the measurement. Moreover, it allows estimating interactions in virtually impossible to demagnetize systems with rectangular major loops.

\section{General $\delta M_{\rm R}$ plot}
Here a $\delta M_{\rm R}$ plot more general than that\cite{dMr-2018} introduced for symmetric hysteresis loops is introduced, allowing to assess interactions even in EB systems. The technique is applied to analyze data obtained at 300\,K via EZ9 MicroSense vibrating sample magnetometer on a magnetron-sputtered, onto a Si(100) substrate, Ta(5\,nm)/Ru(15\,nm)/Co(5\,nm)/IrMn(7\,nm)/Ta(3\,nm) film, where IrMn refers to a (111)-textured Ir$_{20}$Mn$_{80}$ layer, and on a film with the same composition except for it does not contain IrMn. The structural and magnetic properties of these films are reported in Ref.\,\onlinecite{Harres-JAP-2013}. The EB direction of the Co/IrMn film was set by an in-plane magnetic field applied during deposition. Its direction is given by ${\phi_H}$, where ${\phi_H}$$\,=\,$$0^{\circ}$ and $180^{\circ}$ refer to $\mathbf{H_{\rm ext}}$ parallel and antiparallel to the EB direction.
\begin{figure}[t]
\centering
\includegraphics[width=8.2cm]{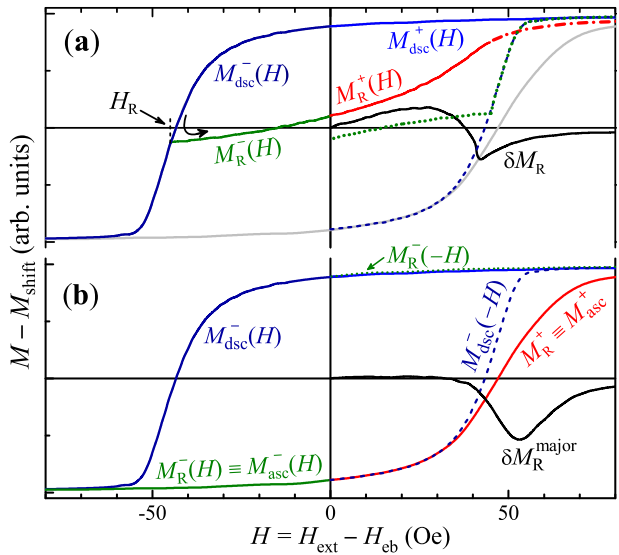}
\caption{Magnetization curves measured for the IrMn/Co film with $H_{\rm eb}\,$$=$$\,-242$\,Oe for ${\phi_H}$$\,=\,$$0^{\circ}$. (a) Representative recoil loop $M_{\rm R}(H)$ and the corresponding $\delta M_{\rm R}$ plot. Here $M_{\rm dsc}^-(-H)$, dash line, and $M_{\rm R}^-(-H)$, dot line, are the curves symmetric through the origin of $M_{\rm dsc}^-(H)$ and $M_{\rm R}^-(H)$. The ascending branch of the major loop (grey line) is also plotted; (b) using the latter as $M_{\rm R}(H)$ gives rise to a nonzero $\delta M_{\rm R}^{\rm major}$.
} \label{centered-dMr}
\end{figure}

The determination of $H_{\rm eb}$ is correlated to that of the coercivity ($H_{\rm c}$) which, normally, is considered as the half-width at half-height of a hysteresis loop. In EB systems, however, due to the characteristic loop's asymmetry, a more general definition\cite{Hsw,Harres-JAP-2013} is used. It employs $H_{\rm sw1}$ and $H_{\rm sw2}$, i.e., the respective switching fields of $M_{\rm dsc}$ and $M_{\rm asc}$, resulting in $H_{\rm c}$$\,=\,$$\frac{1}{2}(H_{\rm sw2}$$\,-\,$$H_{\rm sw1})$ and $H_{\rm eb}$$\,=\,$$\frac{1}{2}(H_{\rm sw1}$$\,+\,$$H_{\rm sw2})$. Here, $H_{\rm sw1}$ and $H_{\rm sw2}$ are the positions of the peaks of the first-order field derivatives of $M_{\rm dsc}^-$ and $M_{\rm dsc}^+$, respectively.

First, a relation interconnecting four parts of a recoil loop allowing the definition of the general $\delta M_{\rm R}$ plot is derived. A $M_{\rm R}(H)$ loop, measured after positive saturation for ${\phi_H}$$\,=\,$$0^{\circ}$ of the Co/IrMn film, is given in Fig.\,\ref{centered-dMr}(a) where $M\,$$-$$\,M_{\rm shift}$ is plotted as a function of $H\,$$=$$\,H_{\rm ext}\,$$-$$\,H_{\rm eb}$, being $M_{\rm shift}$ the shift of the major loop along the magnetization axis (a non-zero $M_{\rm shift}$ could be observed in a system for which, e.g., the maximum negative field ($-H_{\rm max}$) is unable to reverse some positively-saturated magnetization components). Up to $H$ equal to the recoil field $H_{\rm R}$, the descending parts with positive and negative $H$ values of the respective recoil loop coincide with those of the major loop, $M_{\rm dsc}^+(H)$ and $M_{\rm dsc}^-(H)$. The ascending parts of the recoil loop, with negative and positive $H$ are denoted here as $M_{\rm R}^-(H)$ and $M_{\rm R}^+(H)$. A recoil curve is traced after some soft magnetization components have rotated irreversibly along $M_{\rm dsc}^-(H)$. Since for ideal systems only reversible changes occur along $M_{\rm R}^-(H)$, the magnetization varies by $M_{\rm R}^-(H)$$\,-\,$$M_{\rm dsc}^-(H)$ for $H\,$$<$\,0. Along $M_{\rm R}^+(H)$, the soft components reverse back their magnetizations and the respective variation equals $M_{\rm dsc}^+(H)$$\,-\,$$M_{\rm R}^+(H)$. The two variations differ only in sign, so
\begin{eqnarray}
\label{Rec-Rec}
M_{\rm R}^+(H) - M_{\rm dsc}^+(H) = M_{\rm R}^-(H) - M_{\rm dsc}^-(H).
\end{eqnarray}
One can further extend the recoil curve by assuming that, in the $-H_{\rm max}$$\,\leq\,$$H$$\,\leq\,$$H_{\rm R}$ field region, $M_{\rm R}^-(H)$$\,\equiv\,$$M_{\rm dsc}^-(H)$, making Eq.\,\ref{Rec-Rec} valid for all $H$.
Let refer to $M_{\rm dsc}^-(-H)$ as the curve symmetric of $M_{\rm dsc}^-(H)$ through the center of the major loop, and to $M_{\rm R}^-(-H)$ as the curve symmetric of $M_{\rm R}^-(H)$, i.e., $M_{\rm dsc}^-(-H)$$\,=\,$$-M_{\rm dsc}^-(H)$ and $M_{\rm R}^-(-H)$$\,=\,$$-M_{\rm R}^-(H)$. Utilizing these curves (see Fig.\,\ref{centered-dMr}), we define
\begin{eqnarray}\label{dMr}
\delta M_{\rm R}(H) & = & M_{\rm R}^+(H)   + M_{\rm R}^-(-H) \nonumber \\
                    & - & M_{\rm dsc}^+(H) - M_{\rm dsc}^-(-H).
\end{eqnarray}
Note that the ascending part of the major loop does not take the part of the above equations. Evidently, the plot introduced for symmetric loops is a special case of the general $\delta M_{\rm R}(H)$, where $M_{\rm R}(H)$$\,\equiv\,$$M_{\rm asc}(H)$. Besides a recoil loop, the only parameter needed for the construction of a general $\delta M_{\rm R}(H)$ is the position of the center of the major loop ($H_{\rm eb}$,\,$M_{\rm shift}$).

\section{Results and discussions}
For the case of uniaxial anisotropy, nonzero deviations of $\delta M_{\rm R}(H)$ are ascribed to magnetic interactions. The shape of $\delta M_{\rm R}$ shown in Fig.\,\ref{centered-dMr}(a) is similar to that of the plot obtained for the unbiased Co film with symmetric major loop.\cite{dMr-2018} For thin films, an initial increase of $\delta M(H)$ and $\delta M_{\rm R}(H)$ is attributed to parallel (ferromagnetic) exchange coupling and a negative dip to antiparallel (dipolar-like) interactions. However, as it will be demonstrated below, at least part of the negative $\delta M_{\rm R}(H)$ could result from the asymmetry of the magnetization reversal typical for EB systems.

The technique can also be applied to major loops by taking $M_{\rm asc}(H)$ as recoil curve in Eq.\,\ref{dMr}. The asymmetry of an EB major loop, with one of its branches steeper than the other, results in an essentially nonzero $\delta M_{\rm R}^{\rm major}(H)$ as seen in Fig.\,\ref{centered-dMr}(b). Such a plot of the unbiased Co film with symmetric major loop equals zero for any $H_{\rm ext}$. Thus, the $\delta M_{\rm R}^{\rm major}$ plot in Fig.\,\ref{centered-dMr}(b) is a footprint of FM/AF interface coupling.

\begin{figure}[t]
\centering
\includegraphics[width=8.2cm]{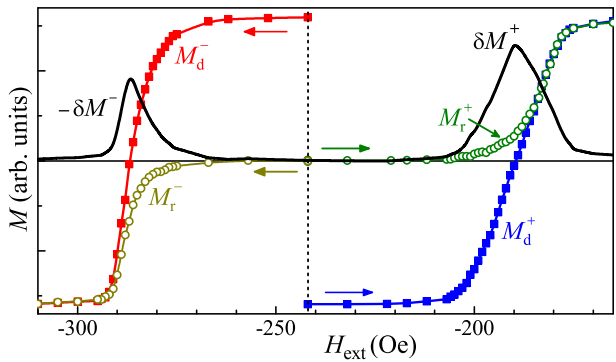}
\caption{Remanence curves for ${\phi_H}$$\,=\,$$0^{\circ}$ and the respective $\delta M$ plots for the Co/IrMn film. Here, $M_{\rm r}^{-}(H_{\rm eb})$$\,=\,$$0$ and $M_{\rm r}^{+}(H_{\rm eb})$$\,=\,$$0$ were attained by dc demagnetization after either positive (left curves) or negative (right curves) saturation. The notations `+' and `$-$' refer to the route of the measurement, e.g., $M_{\rm r}^{-}(H_{\rm ext})$ and $M_{\rm d}^{-}(H_{\rm ext})$ are obtained for $-H_{\rm max}$$\,\leq\,$$H_{\rm ext}$$\,\leq\,$$H_{\rm eb}$, resulting in $\delta M^{-}(H_{\rm ext})$. } \label{Rems}
\end{figure}

The information concerning interactions estimated from the $\delta M_{\rm R}$ plot from Fig.\,\ref{centered-dMr}(a) could be compared to that obtained from the remanence $\delta M$ plots displayed in Fig.\,\ref{Rems}. Since our Co/IrMn film presents EB, these plots [where the states of $M_{\rm r}(H_{\rm eb})$$\,=\,$0 were attained by dc demagnetization] are obtained following the method introduced in Ref.\,\onlinecite{Harres-JAP-2013}. The shapes of the two such plots displayed in Fig.\,\ref{Rems} are characteristics for ferromagnetic coupling with no indication for presence of demagnetizing interactions, differently from $\delta M_{\rm R}(H)$.

This dissimilarity should be attributed to the distinct routines used by the remanence and in-field magnetization techniques. Here, the $\delta M^{-}$ plot is obtained from remanence curves measured for $-H_{\rm max}$$\,\leq\,$$H_{\rm ext}$$\,\leq\,$$H_{\rm eb}$, and $\delta M^{+}$ derives from curves traced for $H_{\rm eb}$$\,\leq\,$$H_{\rm ext}$$\,\leq\,$$H_{\rm max}$. Each plot reflects magnetization reversals that occur along either the descending or the ascending branches of the major hysteresis loop. In contrast, a $\delta M_{\rm R}$ is generated from a recoil loop with $H_{\rm ext}$ cycled following the $H_{\rm max}$$\,\rightarrow\,$$H_{\rm eb}$$\,\rightarrow\,$$H_{\rm R}$$\,\rightarrow\,$$H_{\rm eb}$$\,\rightarrow\,$$H_{\rm max}$ path. According to its definition, $\delta M_{\rm R}$ correlates processes taking place along the descending loop's branch with processes occurring along the ascending branch. Thus, $\delta M_{\rm R}$, differently from $\delta M$, evidences effects steaming from the asymmetry of the reversal, indicating that the negative part of the $\delta M_{\rm R}$ from Fig.\,\ref{centered-dMr}(a) might originate from this asymmetry.

\begin{figure}[t]
\centering
\includegraphics[width=8.5cm]{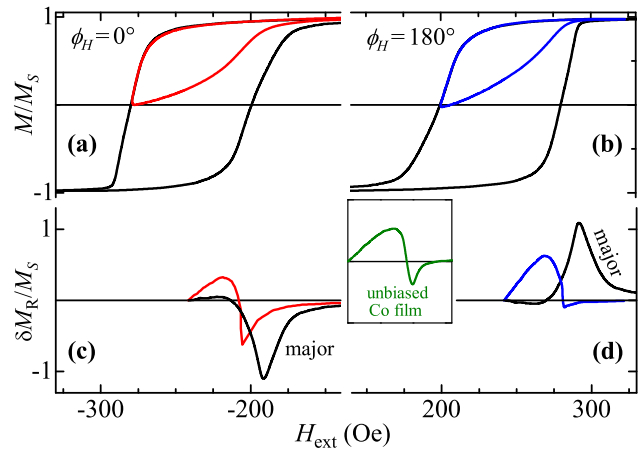}
\caption{Major loops and recoil loops with $H_{\rm R}$$\,\approx\,$$H_{\rm c}$, measured for the Co/IrMn bilayer for ${\phi_H}$$\,=\,$$0^{\circ}$ (a) and $180^{\circ}$ (b). The resultant $\delta M_{\rm R}$ and $\delta M_{\rm R}^{\rm major}$ plots from (a) and (b) are given in (c) and (d), respectively. The $\delta M_{\rm R}$ plot obtained for the unbiased Co film is shown in the inset.} \label{2dMr}
\end{figure}

Major and recoil loops with $H_{\rm R}$$\,\approx\,$$H_{\rm c}$, measured for ${\phi_H}$$\,=\,$$0^{\circ}$ and $180^{\circ}$ for descending fields of the Co/IrMn bilayer, are shown in Figs.\,\ref{2dMr}(a) and (b); the resultant $\delta M_{\rm R}^{\rm major}$ and $\delta M_{\rm R}$ plots are given in Figs.\,\ref{2dMr}(c) and (d). Due to the greater slope of the descending branch as compared to the ascending one of the major loop traced for ${\phi_H}$$\,=\,$$0^{\circ}$, its $\delta M_{\rm R}^{\rm major}$ is virtually negative. Given the reversed asymmetry of the $180^{\circ}$ major loop (with descending branch with lesser slope than the ascending one), the $\delta M_{\rm R,180^{\circ}}^{\rm major}$ is mainly positive. It is identical to $-\delta M_{\rm R,0^{\circ}}^{\rm major}$ though shifted in field by $2|H_{\rm eb}|$.

On the other hand, the asymmetry of the magnetization reversal results in rather different $\delta M_{\rm R}$ plots for ${\phi_H}$$\,=\,$$0^{\circ}$ and $180^{\circ}$. While the former presents an initial increase followed by a negative dip, the latter is essentially positive. Its deviations from the zero line are almost negligible in the field region where the respective $\delta M_{\rm R}^{\rm major}$ initiates its growth. Noteworthy, $\delta M_{\rm R}$ in Fig.\,\ref{2dMr}(c) changes from positive to negative at virtually the same field at which the negative growth of $\delta M_{\rm R}^{\rm major}$ begins. These features strongly support the suggestion that the effects of the reversal's asymmetry determining the shape of $\delta M_{\rm R}^{\rm major}$ are also evidenced in $\delta M_{\rm R}$.

It is instructive to further elucidate the effects on the interaction plots caused exclusively by changes of the FM/AF interface coupling. Our Co/IrMn film presents very stable EB properties given that its $H_{\rm eb}$ and $H_{\rm c}$ values have not practically changed over the six-year period after its deposition. Nevertheless, it is possible to induce significant variations of $H_{\rm eb}$, and even reverse the EB direction, conducting the following experiment. A piece of the bilayer was kept at 300\,K in 4\,kOe static magnetic field, sufficient to saturate the FM along the direction {\it antiparallel} to the EB one. After certain time intervals, hysteresis and recoil loops were measured and immediately after that the sample was placed back at the configuration with ${\phi_H}$$\,=\,$$180^{\circ}$. Temporal changes of $H_{\rm eb}$, i.e., the so-called thermal EB field drift,\cite{Ehresman-jpd-2005} were thus obtained. The drift is ascribed to thermally-activated rotations of some biasing uncompensated spins (UCSs) located at the FM/AF interface away from their initial directions owing to the torque exerted by the negatively-saturated adjacent FM. This has led to a gradual decrease of $H_{\rm eb}$. Considering that at the conditions of this experiment the properties of the Co film have not changed, any modification of its magnetic behavior should be associated with variations of the FM/AF interface coupling only.
\begin{figure}[t]
\centering
\includegraphics[width=8.5cm]{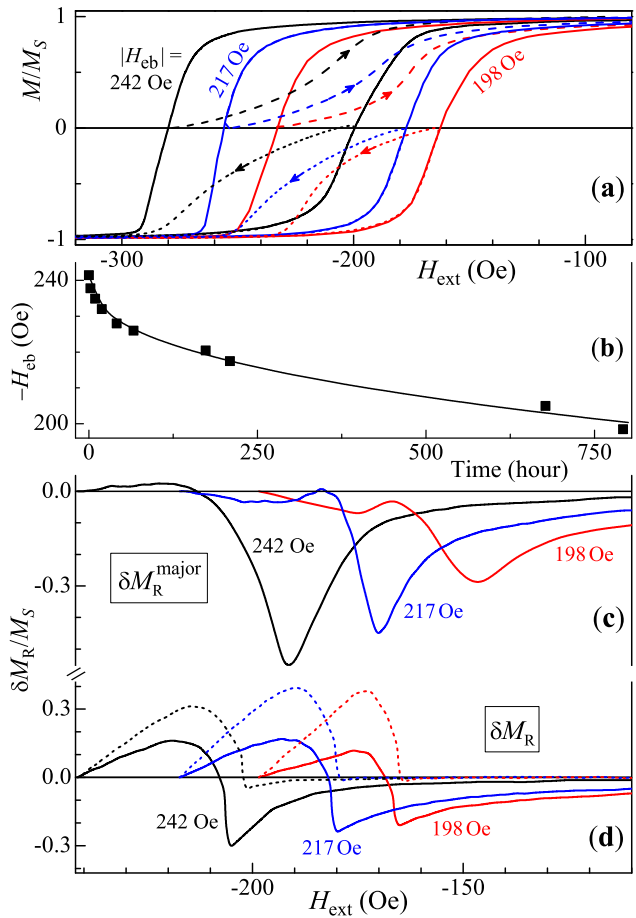}
\caption{(a) Representative easy-axis major hysteresis loops and recoil loops with $H_{\rm R}$$\,\approx\,$$H_{\rm c}$ (dashed lines) of the Co/IrMn film measured after progressively-increasing time intervals with sample kept with saturating $\mathbf{H_{\rm ext}}$ antiparallel to the EB direction; each magnetization is normalized to the respective saturation value $M_s$. (b) Temporal change of $H_{\rm eb}$. The respective $\delta M_{\rm R}^{\rm major}$ and $\delta M_{\rm R}$ are shown in (c) and (d); the plots obtained for ${\phi_H}$$\,=\,$$180^{\circ}$, dash lines in (d), are shifted by $2H_{\rm eb}$.} \label{Tshift}
\end{figure}

Major hysteresis loops with decreasing (due to the above treatment) $H_{\rm eb}$ and the respective recoil loops with $H_{\rm R}$$\,\approx\,$$H_{\rm c}$ measured for ${\phi_H}$$\,=\,$$0^{\circ}$ and $180^{\circ}$ of the Co/IrMn film are given in Fig.\,\ref{Tshift}(a). The EB, see Fig.\,\ref{Tshift}(b), has reduced by almost 20\% after 33 days of such treatment. The corresponding $\delta M_{\rm R}^{\rm major}$ and $\delta M_{\rm R}$ plots are shown in Figs.\,\ref{Tshift}(c) and (d). The former are virtually negative, showing an apparent trend of reduction of their minimum values with the decrease of $H_{\rm eb}$. The $\delta M_{\rm R}$ plots for each field orientation are qualitatively very similar, with nearly identical positive parts. While the negative parts of the $\delta M_{\rm R}$ plots obtained for ${\phi_H}$$\,=\,$$180^{\circ}$ are insubstantial, the $\delta M_{\rm R}$ plots yielded for ${\phi_H}$$\,=\,$$0^{\circ}$ present the same tendency of reduction of the intensity of the minimum as that of $\delta M_{\rm R,0^{\circ}}^{\rm major}$.

The characteristics of the interaction plots in Figs.\,\ref{2dMr} and \ref{Tshift} could be explained as follows. In the absence of Co/IrMn exchange coupling, $\delta M_{\rm R}$ obtained for ${\phi_H}$$\,=\,$$0^{\circ}$ and $180^{\circ}$ with one and the same $H_{\rm R}$ would be identical, presenting an initial rise with maximum value greater than that of the absolute value of the subsequent minimum. Such a plot, obtained for the unbiased Co film, is shown in the inset of Fig.\,\ref{2dMr}. The FM/AF interface coupling could result in shifted major hysteresis loops. In case these are also asymmetric, one obtains nonzero $\delta M_{\rm R}^{\rm major}$. For systems which, for ${\phi_H}$$\,=\,$$0^{\circ}$ are easier to demagnetize than to magnetize given that the descending branch of the major loop is steeper than the other, $\delta M_{\rm R,0^{\circ}}^{\rm major}$ is virtually negative. As already mentioned, the shifted by $2|H_{\rm eb}|$ major plot $\delta M_{\rm R,180^{\circ}}^{\rm major}$ equals $-\delta M_{\rm R,0^{\circ}}^{\rm major}$, so it is mainly positive. The interface coupling should also affect, in a similar manner, the $\delta M_{\rm R}$ plots for ${\phi_H}$$\,=\,$$0^{\circ}$ and $0^{\circ}$ which are altered, in relation to the unbiased one, in just opposite ways. It is reasonable to accept that each $\delta M_{\rm R}$ of our Co/IrMn bilayer is, formally, a superposition of that of the unbiased Co film (given in the inset of Fig.\,\ref{2dMr}) and a curve correlated to the respective $\delta M_{\rm R}^{\rm major}$ though with smaller (most likely proportional to the fraction of $M$ reversed from $H_{\rm eb}$ to $H_{\rm R}$) amplitude.

The positive parts of all $\delta M_{\rm R}$ plots in Fig.\,\ref{Tshift}(d) are not essentially altered by the superposition since $\delta M_{\rm R}^{\rm major}$ is roughly nil in the respective field regions. The negative $\delta M_{\rm R}(H_{\rm ext})$ regions, however, are markedly affected. Those obtained for ${\phi_H}$$\,=\,$$0^{\circ}$ become deeper and the respective field region extends as compared to that of the unbiased film, resembling the characteristics of the (negative) $\delta M_{\rm R,0^{\circ}}^{\rm major}$ plots. The $\delta M_{\rm R}$ plots obtained for ${\phi_H}$$\,=\,$$180^{\circ}$, on the other hand, do not practically retain negative values, eradicated by the superimposed curves proportional to the (positive) $\delta M_{\rm R,180^{\circ}}^{\rm major}$ plots.

Thus, at least for our bilayer, the cross-examination of the major and the pair of recoil-loop plots allowed distinguishing effects coming from magnetic coupling into the FM layer (the initial, positive part of $\delta M_{\rm R}$) from those stemming from interactions at its interface with the AF (the negative $\delta M_{\rm R}$ for ${\phi_H}$$\,=\,$$0^{\circ}$); the proper existence of a non-zero $\delta M_{\rm R}^{\rm major}$ plot is a signature of FM/AF coupling.

\begin{figure}[t]
\centering
\includegraphics[width=8.5cm]{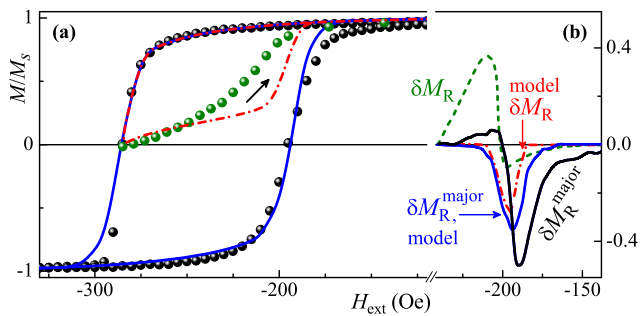}
\caption{(a) Symbols: major hysteresis loop and a recoil loop of the Co/IrMn film measured at ${\phi_H}$$\,=\,$$0^{\circ}$ after it was kept for 1\,hour at ${\phi_H}$$\,=\,$$180^{\circ}$ in $H_{\rm ext}$$\,=\,$4\,kOe. Lines: curves calculated via the polycrystalline model for EB using $t_{set}\,$$=$$\,t_{rot}\,$$=$$\,0.5$\,nm, $m_{set}\,$$=$$\,m_{rot}\,$$=$$\,0.64 M_{FM}\,$$=$$\,900$\,emu/cm$^3$, $J_{set}t_{set}\,$$=$$\,24 $$J_{rot}t_{rot}\,$$=$$\,1.67\,$$\times$$\,10^{-1}$\,erg/cm$^2$, Gaussian FM easy-axis distribution with 3$0^{\circ}$ standard deviation, equally distributed easy axes of the $rot$-type UCSs, and uniaxial anisotropies: $K_{FM}\,$$=$$\,7.35\,$$\times$$\,10^4$\,erg/cm$^3$, $K_{set}\,$=$\,68\,K_{rot}\,$= $\,9\,$$\times$$\,10^6$\,erg/cm$^3$. (b) The respective $\delta M_{\rm R}^{\rm major}$ and $\delta M_{\rm R}$ plots.}
\label{fit}
\end{figure}

Recoil loops and $\delta M_{\rm R}$ plots might provide valuable information on the magnetization reversal mechanism associated to differences in the nucleation process, e.g., these can indicate whether the reversal occurs via either domain-wall motion or coherent rotation.\cite{McCord_2003,Fitzsimmons_2000, Gierlings_2002} Figure\,\ref{fit} shows a pair of experimental major and recoil loops measured for the Co/IrMn film together with fitting curves calculated through the polycrystalline model for EB.\cite{EB-model} It considers that the FM consists of small-sized domains and that at the FM/AF interface there exist grains with UCSs (with magnetization $m$ and thickness $t$) interacting with the FM. These grains, depending on the values of their anisotropy ($K$) and magnetic coupling ($J$, with the adjacent FM) constants, are considered as unstable (i.e., rotatable $rot$, bound to the enhancement of $H_{\rm c}$) or set ($set$, responsible for the bias). At first glance, it might seem that the agreement between experimental and fitting major loops is very reasonable and one may conclude that this model, which considers coherent rotation only, is appropriately chosen. The experimental recoil curve, however, diverges from the model one. Whilst the respective $\delta M_{\rm R}^{\rm major}$ plots are qualitatively similar, the $\delta M_{\rm R}$ plots are very distinct. Although the experimental $\delta M_{\rm R}$ indicates the existence of both intralayer and interlayer interactions, the model plot reflects interface coupling only. I.e., a single $\delta M_{\rm R}$ plot would indicate which types of coupling should be considered to describe the magnetic behavior of the system. For our Co/IrMn film, accounting for exchange coupling (not considered by the model) into the Co layer seems imperative.

Until now, we constructed $\delta M_{\rm R}$ plots from loops with $H_{\rm R}$$\,\approx\,$$H_{\rm c}$ mostly, aiming to compare the method with the remanence one which yields one $\delta M(H)$ only. Obviously, more detailed information could be obtained from a family of $\delta M_{\rm R}$ plots. The potentiality to measure a great number of FORCs and construct a family of $\delta M_{\rm R}$ plots enables (in a manner similar to that FORC diagrams are created) to obtain 3D interaction plot diagrams, 2D diagrams with gradient sets, or interaction effects' distributions plotted inside the hysteresis loops.\cite{FORCs-2007,Khanal-Gallardo,Toro-CM-2017}

\begin{figure}[t]
\centering
\includegraphics[width=8.6cm]{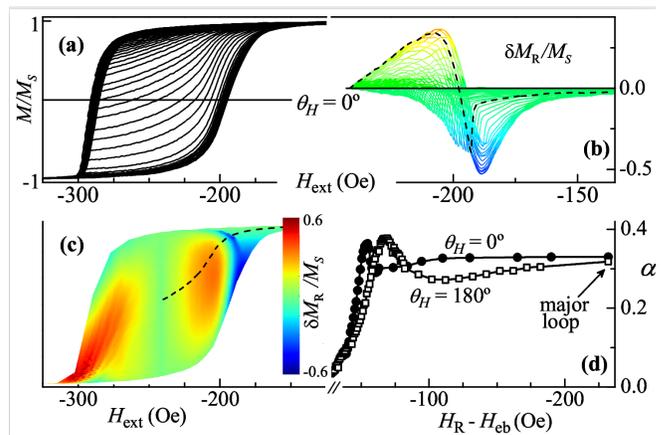}
\caption{FORCs (a) and $\delta M_{\rm R}$ plots (b) of the Co/IrMn bilayer. Map of the $\delta M_{\rm R}(H_{\rm R},H_{\rm ext})$ values plotted inside the major hysteresis loop (c). The right half of the map is obtained using the data from panels (a) and (b), and the left half from data attained for ${\phi_H}$$\,=\,$$180^{\circ}$ (not shown). The $\delta M_{\rm R}$ plot denoted by the dashed line in (b) corresponds to the path shown by a dashed line in (c). The variations of the interaction parameter $\alpha$ with $H_{\rm R}$ for ${\phi_H}$$\,=\,$$0^{\circ}$ and $180^{\circ}$ are given in (d).}
\label{ColorMap}
\end{figure}

Figures \ref{ColorMap}(a) and (b) present a series of FORCs and the respective $\delta M_{\rm R}$ plots for the Co/IrMn film. To each magnetization point of a recoil curve $M_{\rm R}(H_{\rm ext})$ in the $M$$\times$$H_{\rm ext}$ space, one can assign a color associated with the intensity of $\delta M_{\rm R}(H_{\rm ext})$ calculated from this recoil curve. A map of the $\delta M_{\rm R}(H_{\rm R},H_{\rm ext})$ values, plotted inside the major hysteresis loop is given in Fig.\,\ref{ColorMap}(c); the color scale used in (b) is the same as that in (c). While the right half of the map makes use of the data from panels (a) and (b), the left half comes from data attained for ${\phi_H}$$\,=\,$$180^{\circ}$ (not shown). Such a $\delta M_{\rm R}$ diagram depicts how the $\delta M_{\rm R}$ intensity varies along each recoil path. The evocative $\delta M_{\rm R}(H_{\rm ext})$ denoted by a dashed line in Fig.\,\ref{ColorMap}(b) corresponds to the path shown by the dashed line in (c). This $\delta M_{\rm R}$ diagram confirms, e.g., that $\delta M_{\rm R}$ plots obtained at ${\phi_H}$$\,=\,$$180^{\circ}$ on the Co/IrMn bilayer do not show, essentially, negative parts as already seen in Fig.\,\ref{Tshift}(d).

Differently from the FORC method, which provides information on distributions of coercive and interaction fields, a $\delta M_{\rm R}$ diagram insights on interaction effects solely. Surely, analysis of data obtained through the two methods should give a better insight on the interactions present in the system under consideration.

Although in the majority of the studies using remanence plots the estimation of magnetic interactions is purely qualitative, attempts to obtain quantitative information have also been made.\cite{Che-Bertram-1992,Harrell-Alvarez, Toro-CM-2017} In the model of Che and Bertram,\cite{Che-Bertram-1992} the effective interaction field is represented by a mean field and a fluctuation field. The parameter of the mean-field term equals the area under the remanence $\delta M$ vs. $H/H_{\rm r}$ curve (here $H_{\rm r}$ is the remanence coercivity). Likewise, here we define an interaction parameter $\alpha$ as the area enclosed by the $\delta M_{\rm R}/M_s$ vs. $H_{\rm ext}/H_c$ curve. It can also be used as a quantitative measure of the major loop's asymmetry since, for symmetric major loops, $\delta M_{\rm R}^{\rm major}$$\,=\,$0 and so does $\alpha$. Such a parameter should certainly be employed in phenomenological models (yet to be developed) for quantitative assessment of interactions through $\delta M_{\rm R}$ plots. The variations of $\alpha$ with $H_{\rm R}$ for both measurement configurations, ${\phi_H}$$\,=\,$$0^{\circ}$ and $180^{\circ}$, are given in Fig.\,\ref{ColorMap}(d). Whereas due to the reversal asymmetry the two curves are not identical, these present practically one and the same amplitude, though attained at different $H_{\rm R}$ values.

Certainly, more systematic studies on the method should be conducted in a variety of systems to clarify its full potentiality for interaction effects estimations.

\section{Summary and conclusions}
The general $\delta M_{\rm R}$ plot introduced here and applied to Co/IrMn gives a simple, yet efficient, way to assess interactions even in systems with shifted and asymmetric major hysteresis loops. The essentially nonzero $\delta M_{\rm R}^{\rm major}$ plot revealed to be a precise measure of the reversal asymmetry. Also, the two distinct $\delta M_{\rm R}$ plots, obtained for the same recoil field but for measuring field parallel or antiparallel to the exchange-bias direction, together with the two-dimensional $\delta M_{\rm R}$ diagram, provide valuable information on the magnetization reversal mechanism and allow distinguishing effects coming from magnetic coupling into the ferromagnet from those stemming from interactions in its interface with the antiferromagnet. Also, the here-defined interaction parameter, i.e., the area enclosed by a $\delta M_{\rm R}$ curve, could be used to quantitative measure the interaction effects and the major loop's asymmetry. This technique can readily be adjusted for assessing effects caused by deviations from theoretical behavior of other hysteretic quantities.

Free software which generates the introduced here general $\delta M_{\rm R}$ plots and also yields $\alpha$ is available for download at http://www.if.ufrgs.br/pes/lam/dMr.html.

\section{Acknowledgments}
The research has been partly financed by the Brazilian agencies CNPq (grants 305796/2016-0 and 422740/2018-7) and CAPES (finance code 001).

\end{document}